# The Origin of Fe II Emission in AGN

J.A. Baldwin

*Physics and Astronomy Department, Michigan State University, 3270 Biomedical Physical Sciences Building, East Lansing, MI 48824*

G.J. Ferland

*Department of Physics and Astronomy, University of Kentucky, 177 Chemistry/Physics Building, Lexington, KY 40506*

K.T. Korista

*Department of Physics, University of Western Michigan, 1120 Everett Tower, Kalamazoo, MI 49008-5252*

F. Hamann

*Department of Astronomy, University of Florida, 211 Bryant Space Center, Gainesville, FL 32611-2055*

A. LaCluyzé

*Physics and Astronomy Department, Michigan State University, 3270 Biomedical Physical Sciences Building, East Lansing, MI 48824*

baldwin@pa.msu.edu

## Abstract

We used a very large set of models of broad emission line (BEL) clouds in AGN to investigate the formation of the observed Fe II emission lines. We show that photoionized BEL clouds cannot produce both the observed shape and observed equivalent width of the 2200 – 2800Å Fe II UV bump unless there is considerable velocity structure corresponding to a microturbulent velocity parameter $v_{turb} \geq 100$ km s$^{-1}$ for the LOC models used here. This could be either microturbulence in gas that is confined by some phenomenon such as MHD waves, or a velocity shear such as in the various models of winds flowing off the surfaces of accretion disks. The alternative way that we can find to simultaneously match both the observed shape and equivalent width of the Fe II UV bump is for the Fe II emission to be the result of collisional excitation in a warm, dense gas. Such gas would emit very few lines other than Fe II. However, since the collisionally excited gas would constitute yet another component in an already complicated picture of the BELR, we prefer the model involving turbulence. In either model, the strength of Fe II emission relative to the emission lines of other ions such as Mg II depends as much on other parameters (either $v_{turb}$ or the surface area of the collisionally excited gas) as it does on the iron abundance. Therefore, the measurement of the iron abundance from the FeII emission in quasars becomes a more difficult problem.

*Subject headings:* quasars: emission lines, galaxies: active

# 1. Introduction: The Importance of Fe II Emission from AGN

Fe II is an important contributor to the emission line spectrum of many AGN. Because of its complex atomic structure, the $Fe^+$ ion emits through a huge number of multiplets scattered across the UV-optical part of the spectrum.

Understanding the physics of Fe II emission is important for several reasons. The emission can be strong, showing that it helps determine the energy budget of the emitting gas, and a measurement of its abundance as a function of cosmic time could verify some cosmological parameters. In most models of the evolution of galaxies, much of the iron in the ISM comes from Type Ia supernovae, which start to occur only about 0.3 to 1 billion years after the onset of star formation. Hence we expect to see a sudden jump in that abundance at some time that we would equate to about 0.3 to 1 billion years after the initial burst of star formation (see Hamann & Ferland 1999; Matteucci & Recchi 2001). Recent detection of the Gunn-Peterson effect (Djorgovski et al. 2001; Becker et al. 2001) indicates that re-ionization of the IGM was finishing at $z = 6$. Thus the major onset of star formation might have occurred as recently as $z \sim 6$, or might have ramped up slowly over the ~half Gyr period between $z \sim 10$ and $z \sim 6$, when the QSOs are hidden from view by H I absorption (e.g. Di Matteo et al. 2003). A properly calibrated iron chronometer would let us test these scenarios through observations of $z = 5$ quasars, offering an independent check on the recent WMAP results indicating that the first stars formed at around $z = 20$ (Bennett et al. 2003).

Primarily because of the connection to early chemical evolution, there has been considerable interest in using near-infrared spectroscopy to measure the strength of Fe II emission in high-redshift QSOs out to $z \sim 6$ (Hill, Thompson & Elston 1993; Dietrich et al. 2002; Iwamuro et al. 2002; Freudling, Corbin, & Korista 2003). The general hope has been to use the ratio of the intensity of UV Fe II lines to that of Mg II λ2800 to obtain the Fe/Mg abundance ratio, in order to see how much material has been processed through Type Ia supernovae vs. how much has gone through the α-process in stars and released via Type II supernove. However, the observations have proceeded in the absence of any calibration to convert the observed relative line strengths into actual abundance ratios. While the mere presence or absence of Fe II emission as a function of lookback time does in itself carry some significance, it is not the same as being able to reliably distinguish between an epoch of rapid iron enrichment as opposed to some change in the average excitation conditions that determine how strongly Fe II lines are emitted.

In the present paper, we investigate in some detail whether or not the Fe abundance in distant QSOs can really be deduced from observations of the Fe II emission lines. We concentrate on the Fe II "UV bump", a broad blend of UV emission that can be measured in objects over a wide range of redshifts.

# 2. Observed Spectra

For comparison with the models that we develop here, Figure 1 shows the previously observed spectra of some AGN with strong Fe II emission.

In high-redshift QSOs, the most easily measured Fe II parameters are the shape and strength of the pronounced bump between the C III] 1909 and Mg II 2800 emission lines (henceforth the "UV bump"), and we will concentrate on those parameters in this paper. The UV bump together with the Balmer continuum emission are often referred to as the "Small Bump". Figure 1 illustrates that the general shape of the UV bump is usually about the same in different AGN, even though the Fe II strength relative to the continuum and the widths of the individual emission lines may be very different from object to object. Our goal is to reproduce both the shape and strength of this feature.

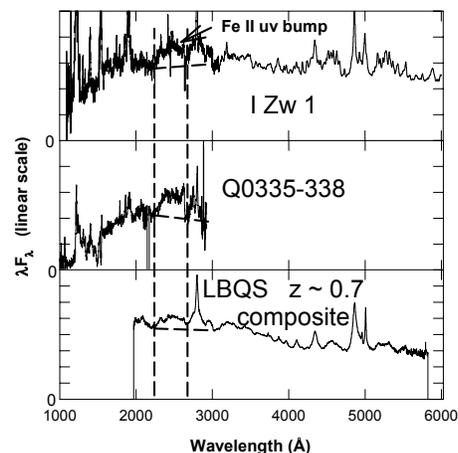

**Figure 1.** Observed spectra, showing how the UV bump is defined. The EW of the UV bump is defined as the Fe II flux above a pseudocontinuum as sketched in the figure, relative to the continuum flux at 2400Å.

I Zw 1 is a Seyfert galaxy often used as a benchmark for Fe II measurements, because of its combination of strong Fe II emission and narrow (FWHM ~ 1500 km s$^{-1}$) emission lines. We combined the ultraviolet HST spectra of I Zw 1 taken by Laor et al (1997) with a ground-based spectrum kindly made available to us (and previously to Laor et al.) by Dr. B.J. Wills, to produce the combined spectrum shown in the upper panel of Figure 1. Although the UV and optical spectra were taken at different times, their continuum levels agree almost exactly, so we combined them without any allowance for possible variations in the emission-line strengths. Vestergaard and Wilkes (2001) have used these same data for a very comprehensive study of the observed Fe II spectrum[1].

The lower two panels of Figure 1 show higher-luminosity examples. QSO 0335-338 is relatively narrow-lined (FWHM = 4000 km s$^{-1}$), with a strong Fe II UV bump (Baldwin et al. 1996). The composite spectrum in the lower panel is assembled from spectra of 15 z ~ 0.7 QSOs in the LBQS survey (Francis et al. 1991), each chosen to have the major UV and optical Fe II emission obtainable in the *same* spectrum.

At lower redshifts, the Fe II bands seen in the Hγ – Hβ region in the I Zw 1 spectrum are also important diagnostics, including their strength relative to the UV bump. In the observed spectra, these optical wavelength Fe II bands carry only about 25–35 percent of the total energy emitted in Fe II (Wills, Netzer & Wills 1985; Dietrich et al. 2002).

The first three rows of Table 2 show measured results for these real AGN, for several parameters that will be described below. In the following sections, we smooth our model spectra to 1500 km s$^{-1}$ FWHM for easy comparison to the spectrum of I Zw 1.

## 3. Modeling Fe II Emission

The observed Fe II spectrum is produced by a variety of processes, including collisional excitation, line trapping and thermalization, and selective excitation by overlaps with other lines. The gas temperature, electron density, and line optical depths all affect the emission. A prediction of the Fe II spectrum requires that all of these effects must be included self-consistently with the temperature and ionization structure of the gas. Our methods are described in Verner et al. (1999).

There have been several previous studies of Fe II emission from AGN. Netzer and Wills (1983; Wills, Netzer & Wills 1985) modeled Fe II emission from QSOs using an approach very similar to ours, but with only a 70-level model of the iron atom. This gave only a very approximate idea of the actual details of the iron emission.

We use a 371-level Fe$^+$ model, including all energy levels up to 11.6 eV and which calculates strengths for 68,000 emission lines (Verner et al. 1999), incorporated into the development version of the *Cloudy* photoionization simulation code (last described by Ferland et al. 1998). Our work complements that of Sigut & Pradhan (2003; see also Sigut, Pradhan & Nahar 2004), who used a more detailed model of the Fe$^+$ ion than we did, with 827 levels and 23,000 lines, and performed an exact calculation of line radiative transfer, but did not use energy balance to obtain a self-consistent temperature and ionization structure. Our models properly calculate a fully self-consistent temperature structure and energy balance. However, the radiative transfer is calculated using the escape probability formalism, which although it is known to reproduce exact results when the source function is constant (Elitzur 1992), is more approximate when conditions vary as in a realistic calculation. The "correct" calculation, one with both exact radiative transport and energy transport, has not been done yet.

---

[1] The spectrum of I Zw 1 shown in Figure 1 includes the 9% correction to the flux measured in the 1087-1606Å wavelength range that is described by Vestergaard and Wilkes.

We can get an idea of the errors introduced by the approximations that are needed to calculate these models on today's computers by comparing our predictions, with approximate radiative transfer but exact energy balance, with the Sigut & Pradhan calculation, with exact radiative transfer but no energy balance. Figure 2 shows this comparison for one of the Sigut & Pradhan cloud models[2]. It is seen that the general agreement is reasonably good, in spite of the very different computational approaches. The differences that are present will not affect our conclusions.

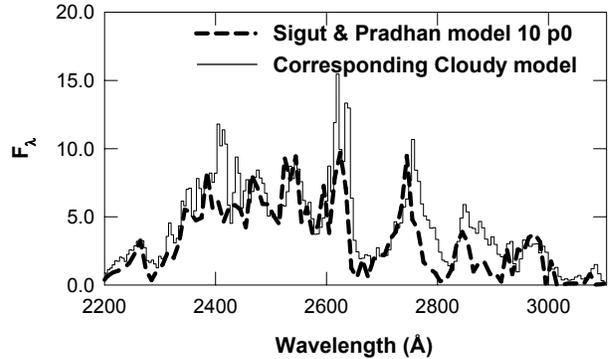

**Figure 2.** Comparison of model 10 p0 from Sigut & Pradhan (2003) to the corresponding Cloudy model.

The limited number of levels in our model atom is also a possible source of error. The highest level is 11.59 eV above ground. Direct collisions to this level from the ground state must be unimportant since the excitation potential is so much greater than the thermal energy. Continuum photoexcitation of FUV transitions is more of a concern since a strong UV continuum is present in quasars. Verner et al. (2003) have recently published results from a model with 830 levels. Their Figure 1 shows that the results are quite similar to those from our 371 level atom, except for a 40% increase in the equivalent width of the Fe II UV bump when the larger model atom is used. The large photon to electron ratio used for this particular calculation results in most of the energy in Fe II being emitted via fluorescence, where the larger atom's impact is most significant. However, it is important to note that these parameters also result in Fe II emission that is far weaker than observed, because the assumed ionization parameter (log $U_H$ = 0.523) is far too large despite the large assumed cloud column density ($N_H$ = $10^{24}$ cm$^{-2}$). It is generally the case that models in which additional levels have a large effect on the predicted spectrum are only very weak Fe II emitters. We will return to this comparison in §6. Our general conclusion is that over the parts of parameter space that are important for producing Fe II emission, our 371-level atom gives sufficient accuracy. The great advantage of using this smaller model atom is that our calculations run rapidly enough that we can explore a very wide range of input parameters.

Any successful model for the formation of Fe II lines in AGN must also be consistent with the data for other emission lines and with the observed equivalent widths. This should also include consistency with the reverberation results that show that the photoionized BELR gas is distributed over a wide radial extent (as is directly observed for many low-luminosity AGN through reverberation observations; e.g. Peterson 1993). The calculations presented below are done in the context of the LOC model (Baldwin et al. 1995) in which clouds exist at a broad range of density and distance from the source of ionization radiation. In this picture of the BELR, selection effects, largely introduced by the atomic physics, ensure that the strongest lines are produced with the observed strengths. This model is known to reproduce the observed intensity ratios of lines other than FeII and broadly reproduces the reverberation results (Korista & Goad 2000). Our standard calculation involves a set of 225 single-cloud models over a grid of points in gas density ($10^7$ cm$^{-3}$ ≤ $n_H$ ≤ $10^{14}$ cm$^{-3}$) and ionizing photon flux ($10^{17}$ cm$^{-2}$ s$^{-1}$ ≤ $\Phi$ ≤ $10^{24}$ cm$^{-2}$ s$^{-1}$). We thus are able to investigate the effects of a range of microturbulent velocity, column density, chemical abundances, etc., as well as experiment with different weighting functions for the numbers of clouds at different points on the $\Phi - n_H$ plane. This lets us clearly identify which assumptions affect the spectrum, and in what ways.

---

[2] The parameters used by Sigut & Pradhan (2003) are: log $n_H$ = 11.6, log $U_H$ = -2 (Log $\Phi$ = 20.08), $v_{turb}$ = 10 km s$^{-1}$, solar abundances but with [Fe/H] = -4.52, and the spectral energy distribution from Mathews and Ferland (1987).

The Fe II spectrum is far too complex for individual lines to be considered, since a particular spectral feature can be the sum of many overlapping lines. Further, quasar emission lines are velocity broadened by an amount between $10^3 - 10^4$ km s$^{-1}$. To compare theory with observations model Fe II spectra are computed for each point on the $\Phi - n_H$ plane by sorting the 68,000 computed Fe II emission line strengths into 1000 wavelength bins, each 580 km s$^{-1}$ wide, centered at gradually increasing wavelength spacings between 1001Å and 6993Å. For comparison to I Zw 1, the model spectra are then convolved with a gaussian smoothing function which has FWHM = 1500 km s$^{-1}$.

## 4. The Baseline Model and its Problems

Our starting point is a series of models with the same parameters as the standard models calculated by Korista et al. (1997) for their atlas of AGN emission line strengths. These include a column density $N_H = 10^{23}$ cm$^{-2}$, a standard AGN ionizing continuum shape (the "baseline SED" presented in Korista et al. 1997), and solar abundances defined by the following logarithmic gas phase chemical abundances relative to hydrogen: H : 0.0000; He: -1.0000; Li: -8.6904; Be:-10.5800; B : -9.1198; C : -3.4498; N : -4.0301; O : -3.1302; F : -7.5200; Ne: -3.9318; Na: -5.6861; Mg: -4.4202; Al: -5.5302; Si: -4.4498; P : -6.4283; S : -4.7905; Cl: -6.7258; Ar: -5.4001; K : -6.8697; Ca: -5.6402; Sc: -8.8013; Ti: -6.9586; V :

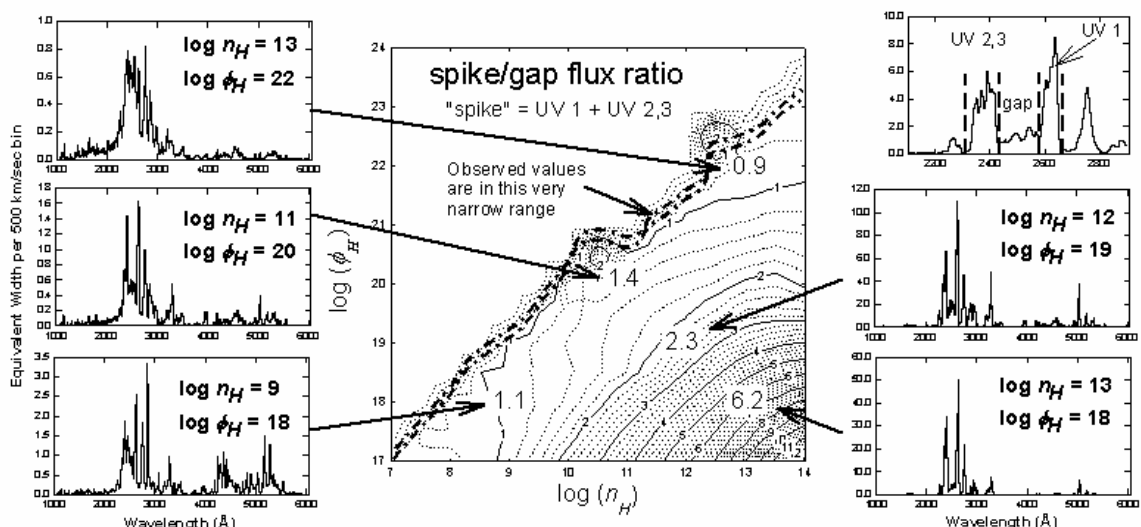

**Figure 3.** Contour plot of the "spike/gap" ratio, which parameterizes the strength of the UV 1,2 and 3 multiplets relative to the rest of the UV bump, as indicated in the upper right box. Sample computed spectra at five locations on the density, ionizing flux ($n_H$, $\Phi_H$) plane are also shown. The large numbers on the contour plot in this (and the similar figures which follow) show the values of the contoured parameter at the positions for which spectra are plotted in the small boxes.

-7.9788; Cr: -6.3152; Mn: -6.4660; Fe: -4.4895; Co: -7.0799;.Ni: -5.7545; Cu: -7.7282; and Zn: -7.3449. For the sake of making consistent comparisons to our previous work, we have *not* included the recent revisions by Allende Prieto et al. (2001, 2002) and Holweger (2001) to the solar O, C and Fe abundances. This calculation assumes thermal line broadening.

### 4.1 Shape of the Fe II UV Bump

Figure 3 shows results from model calculations over the log $\Phi$ – log $n_H$ plane. It is seen from the sample computed spectra plotted in the figure that over much of the parameter space, the computed spectrum is dominated by two strong spikes at about 2400Å and 2600Å, which are not present in the observed spectrum. We will refer to these two features as "the spikes". The spikes consist of the strongest UV transitions, UV 1 (2600Å) and the UV 2+3 (2400Å) multiplets. It is no surprise that models predict these to be the strongest transitions – the puzzling thing is that they are not prominent in AGN. As we shall see, this will sharply constrain the possible models.

The upper right corner of Figure 3 shows how we defined a "spike/gap" parameter by adding up the total computed Fe II flux over the wavelength ranges 2312–2428Å and 2565–2665Å (which encompass the two spikes), and then dividing by the total Fe II flux in the range 2462–2530Å (the region between the two spikes). A contour map of this quantity is shown in the central part of Figure 3.

In the observed spectra, the spike/gap ratio is always very near 0.7, corresponding to the narrow band between the heavy dashed lines cutting across the contour plot from bottom left to top right. In this baseline model, only a tiny fraction of the ($\Phi$, $n_H$) parameter space has the observed spike/gap ratio. One theme of the remainder of this paper is to try to understand why the UV 1, 2 and 3 multiplets are not nearly as strong in the real AGN as in our baseline model spectra.

## 4.2 Fe II Equivalent Width

Figure 4 shows how the equivalent width of the Fe II UV bump depends on $\Phi$ and $n_H$. This is predicted in the same way it was measured in the observed AGN spectra listed in Table 1; a pseudo-continuum (in this case the average of the fluxes at 2001Å and 3052Å) is first subtracted off, and then the integrated Fe II flux above this continuum is added up over the wavelength range 2240–2650Å. This is then converted into the equivalent width of the UV bump by dividing by the incident continuum at 2400Å. The standard continuum shape used in most of our models is similar to the observed shape for the LBQS z ~ 0.7 composite spectrum shown in Figure 1. The heavy dashed line in Figure 4 shows the observed equivalent width (EW) of the UV bump for the LBQS z ~ 0.7 composite spectrum. This is a lower limit on the contour plot because the equivalent widths are plotted for a covering factor of 1.0, while the covering factor in the actual QSOs may be much smaller. Therefore, the whole lower-right portion of Figure 4 is consistent with the observations. However, the UV bump emitted by gas in this region would *not* have the correct shape, as can be seen from Figure 3, while the parameters where the shape is correct produce emission that is far too faint.

## 4.3 Comparison of integrated model spectra to observations

The true BELR is known to have gas at a range of different distances from the continuum source and also with a range of different densities. We represented its integrated spectrum by summing the spectra emitted by individual clouds scattered over the $\log(\Phi) - \log(n_H)$ plane. We integrated over this plane using the standard limits described above and the covering factor dependences $f(r) \propto r^{-1}$ and $g(n_H) \propto n_H^{-1}$. This corresponds to the standard LOC model described by Baldwin et al. (1995).

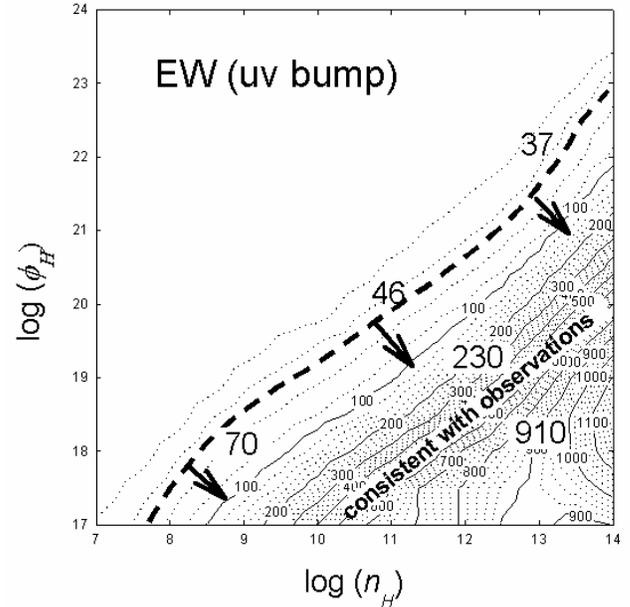

**Figure 4.** Contour plot of equivalent widths calculated for standard photoionization model. The large numbers indicate the equivalent widths at the ($n_H$, $\Phi_H$) points corresponding to the positions of the spectra shown along the sides of Fig. 3. Only the region below the heavy dotted line is consistent with the observations, even for 100% covering factor. This does not overlap with the region in Fig. 3 that is consistent with the observations.

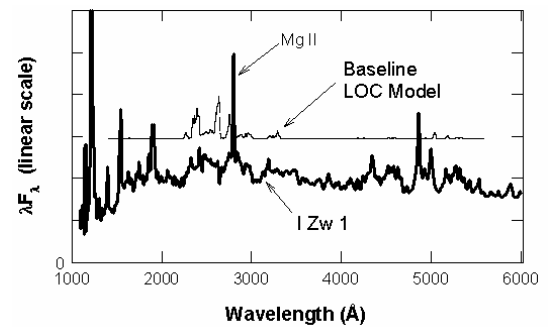

**Figure 5**. Observed spectrum of I Zw 1 (heavy solid line) compared to the Fe II spectrum calculated for our baseline LOC model (thin dashed line). The LOC model is on the same scale as the observed spectrum, but has been shifted upwards by one vertical tickmark. Note the strong spikes in the computed spectrum at about 2400Å and 2600Å, which clearly are not present in the observed spectrum.

| | Log $N_H$ | O VI 1035 + Ly$\beta$ | Ly$\alpha$ | N V 1240 | C IV 1549 | He II 1640 + O III] 1663 | C III] + Si III] + Al III] 1900 | Mg II 2798 | H$\beta$ |
|---|---|---|---|---|---|---|---|---|---|
| **Observed** | | 0.1-0.3 | 1.00 | 0.1-0.3 | 0.4-0.6 | 0.1-0.2 | 0.15-0.3 | 0.15-0.3 | 0.07-0.2 |
| **Models** | | | | | | | | | |
| (1) Baseline log $N_H$ = 23 | 23 | 0.32 | 1.00 | 0.06 | 0.51 | 0.16 | 0.10 | 0.31 | 0.07 |
| (2) Baseline log $N_H$ = 24 | 24 | 0.30 | 1.00 | 0.05 | 0.48 | 0.16 | 0.12 | 0.29 | 0.06 |
| (3) Baseline log $N_H$ = 25 | 25 | 0.18 | 1.00 | 0.04 | 0.49 | 0.15 | 0.11 | 0.36 | 0.12 |
| (4) CNOFe enhanced | 24 | 0.27 | 1.00 | 0.07 | 0.44 | 0.19 | 0.28 | 0.35 | 0.06 |
| (5) [Fe/O] = -0.8 | 24 | 0.30 | 1.00 | 0.05 | 0.48 | 0.16 | 0.12 | 0.32 | 0.06 |
| (6) [CNO/H] = -0.8, [Fe/O] = -0.8 | 24 | 0.22 | 1.00 | 0.03 | 0.40 | 0.13 | 0.03 | 0.14 | 0.06 |
| (7) $\nu^{-1}$ power law continuum | 24 | 0.76 | 1.00 | 0.14 | 1.00 | 0.25 | 0.25 | 0.53 | 0.08 |
| (8) $v_{turb}$ = 100 km s$^{-1}$ | 23 | 0.25 | 1.00 | 0.06 | 0.49 | 0.11 | 0.06 | 0.46 | 0.08 |
| (9) $v_{turb}$ = 1000 km s$^{-1}$ | 23 | 0.17 | 1.00 | 0.05 | 0.36 | 0.09 | 0.06 | 0.46 | 0.08 |

Table 1
Relative Strengths of Strong Lines

Table 2
Observed vs. Model Fe II Parameters
(measured in the same way in all cases)

| | Spike/gap ratio | Covering Factor | EW (UV bump) | F (UV bump)/F (Mg II) |
|---|---|---|---|---|
| **Observed:** | | | | |
| I Zw 1 | 0.7 | --- | 84 | 3.3 |
| Q0335-338 | 0.7 | --- | 130 | 9 |
| LBQS z ~ 0.7 composite | 0.7 | --- | 49 | 2.2 |
| **Models:** | | | | |
| (1) Baseline log $N_H$ = 23 | 2.9 | 0.30 | 58 | 0.70 |
| (2) Baseline log $N_H$ = 24 | 2.7 | 0.29 | 62 | 0.78 |
| (3) Baseline log $N_H$ = 25 | 2.8 | 0.21 | 71 | 0.76 |
| (4) CNOFe enhanced | 2.0 | 0.34 | 175 | 1.92 |
| (5) [Fe/O] = -0.8 | 2.1 | 0.28 | 20 | 0.24 |
| (6) [CNO/H] = -0.8, [Fe/O] = -0.8 | 1.3 | 0.26 | 8 | 0.19 |
| (7) $\nu^{-1}$ power law continuum | 2.4 | 0.25 | 145 | 1.19 |
| (8) $v_{turb}$ = 100 km s$^{-1}$ | 1.0 | 0.18 | 101 | 0.91 |
| (9) $v_{turb}$ = 1000 km s$^{-1}$ | 0.8 | 0.10 | 178 | 1.63 |

Table 1 compares the relative strengths of the usual strong AGN lines with observed values taken from Baldwin et al. (1995). The LOC model constructed from our baseline set of Cloudy models is called Model 1 in the table. The other models will be described later in the paper. The baseline LOC model is a good fit to the observations (it is essentially the same model used by Baldwin et al. 1995 when the LOC model was first proposed).

Table 2 lists the spike/gap ratio, EW (UV bump), and also the F(UV bump)/F(Mg II) flux ratios observed in the three sample observed spectra from §2, and for various LOC models including our baseline model. The covering factor used to calculate the EW (UV bump) in Table 2 for the baseline model, and also for further models described below, is calculated by requiring the sum of the Ly$\alpha$ and Mg II 2800 fluxes, expressed as an equivalent width relative to the continuum at 1216Å, to be 130Å. This is the upper end of the observed range (Baldwin et al. 1995). The baseline LOC model clearly does *not* match the Fe II observations. Figure 5 illustrates how different the Fe II spectrum computed for the baseline LOC model is from the spectrum observed in I Zw 1.

We have used the standard LOC parameters from Baldwin et al. (1995) because they can satisfactorily reproduce the relative strengths of the strong emission lines from elements other than Fe, but Figures 3 and 4 clearly show that models integrated over a more restricted range or with different weightings would not produce Fe II emission matching the observations either, since there is no point on the log $\Phi$ – log $n_H$ plane where any individual model matches the observations.

## 5. Dependence on other cloud properties

Having shown that the standard BELR cloud does not produce the observed Fe II emission, we now examine how the spectrum depends on other parameters such as column density and abundance.

### 5.1 Column density effects.

First we investigate the consequences of column densities greater than those assumed in the standard calculations already presented. Increasing the column density of the individual clouds does change the details of which regions of the log $\Phi$ – log $n_H$ plane produces the spikes. However, even for a very high column density of $N_H = 10^{25}$ cm$^{-3}$, for which the cloud is becoming Compton thick, only two very small regions of the log $\Phi$ – log $n_H$ plane can simultaneously produce the correct shape and equivalent width for the UV bump (Fig. 6, which is for model 3 listed in Tables 1 and 2). However, neither of these regions will produce lines of other elements in the observed intensity ratios and equivalent widths, so it is necessary to integrate over some significant area on the log $\Phi$ – log $n_H$ plane. For example, single models in the lower-left corner of the log $\Phi$ – log $n_H$ plane produce too much Mg II 2800 relative to the Fe II UV bump, while in the allowed region in the upper-right corner of the plot Mg II is much too weak while the C IV/Ly$\alpha$ intensity ratio is far higher than is observed.

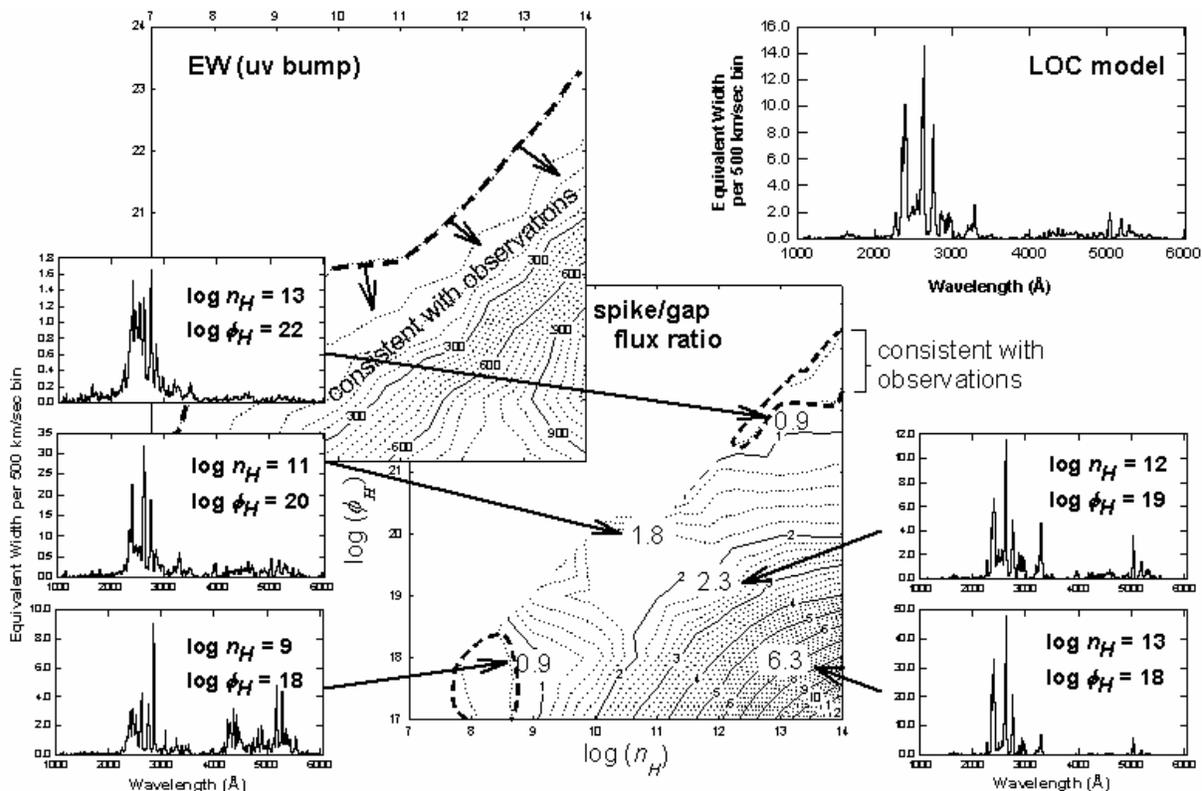

**Figure 6.** Results from a coarse grid of models with column density $N_H = 10^{25}$ cm$^{-3}$. The large numbers, arrows, etc. have the same meanings as in Figs. 3 and 4. In the plot of the spike/gap flux ratio, we exclude contours in regions where the calculations indicate completely negligible emission in Fe II, and the ranges consistent with observation are therefore just the small areas that are shown enclosed in heavy dashed lines, at the bottom-left and top-right of the spike/gap flux ratio plot.

A similar situation applies to the models presented by Verner et al. (2003), which have log $\Phi$ = 20.5, log $n_H$ = 9.5, $v_{turb}$ = 1 km s$^{-1}$, solar abundances and log $N_H$ = 24. The computed shape of the Fe II bump and its intensity ratio relative to Mg II 2800 are in reasonable agreement with the observations, but when we ran the same model using our 371 level Fe II atom (as opposed to the 830 level atom used by Verner et al.), the predicted EW(Fe II UV bump) is only 11Å for a covering factor of 1. As mentioned above, this cloud is too highly ionized to emit much Fe II. Similarly, the Mg II/Ly$\alpha$ intensity ratio predicted by our version of this model is only 0.02, an order of magnitude lower than the observed ratio.

We conclude that a LOC-type integration is required to produce reasonable intensity ratios for the strong lines for these high column-density cases, but that for any such model covering any reasonable range on the log $\Phi$ – log $n_H$ plane, the emission will be dominated by the UV 1 and UV 2+3 spikes.

## 5.2 Chemical abundances and ionizing continuum shape.

We made a series of calculations with modified chemical abundances, to see if such changes would affect the emitted Fe II spectrum in unforeseen ways. LOC integrations over the log $\Phi$ – log $n_H$ plane gave the results listed as Models 4–7 in Tables 1 and 2. These all had $N_H$ = 10$^{24}$ cm$^{-2}$ (as compared to 10$^{23}$ cm$^{-2}$ in the baseline model), but for comparison we included Model 2 which has parameters otherwise identical to those of the baseline model. Model 4 has abundances of all elements heavier than He enhanced by 0.6 dex, except that Fe/O was enhanced by a further factor of 0.5 dex (this represents the end point of the chemical evolution of the giant elliptical galaxy model shown in Fig. 2 of Hamann 1999). Model 5 has Fe/H decreased by 0.8 dex (this represent the giant elliptical model at an age of 0.1 Gyr, before Fe enhancement through Type Ia supernovae). Model 6 has C/H, N/H, O/H and Fe/O all decreased by 0.8 dex (representing a very early stage of the giant elliptical's chemical evolution). These models all used the same ionizing continuum shape as our baseline model.

We also tried (Model 7) a radical modification of the shape of the ionizing continuum spectrum, using a bare power law with $f_\nu \propto \nu^{-1}$, except for an infrared cutoff at 1 μm. This is in contrast to our baseline model, in which the ionizing continuum shape contains a Big Blue Bump (see Korista et al. 1997 for details).

Table 1 shows that model 7, with the hard continuum, does not emit even the non-Fe lines in the observed ratios. However, all of the models in the sequence of different chemical abundances emit the strong lines of elements other than Fe in ratios that are compatible with the observations. We must look at Table 2 to see if these models are viable. As expected, the Fe II/Mg II intensity ratio scales with Fe/O, and high Fe abundances lead to larger equivalent widths for the Fe II UV bump. However, the spike/gap ratio gets higher for higher Fe abundances. Only Model 6, with [CNO/H] = -0.8 and [Fe/O] = -0.8 produces a UV bump that resembles the observed shape, but the equivalent width of the UV bump is only 8Å after accounting for the covering factor needed to keep Ly$\alpha$ and Mg II from being stronger than is observed. There are no models that can simultaneously reproduce all of the observed parameters.

## 6. Photoionized Models with Microturbulence.

The models we have presented so far either produce too little Fe II emission or emission that is dominated by the first three UV multiplets. The basic problem is that photoionization deposits most energy within a few continuum optical depths of the surface. That energy is then mainly radiated through the strongest resonance lines. The observed Fe II emission is so strong that it represents a significant challenge to any photoionized model (see the discussion by Netzer 1985). The Fe II equivalent width problem was first pointed out by Netzer & Wills (1983). They showed that adding microturbulence improves the situation. The increase in equivalent width of FeII with increasing turbulence is due to two effects - larger line widths allow line photons to escape more easily, and the importance of continuum pumping increases since each driving line absorbs mainly photons over one line width. Here we show that microturbulence also helps solve the shape problem discussed above.

Bottorff et al. (2000) have considered the effects of microturbulence on BELR spectra and give computational details, but do not discuss Fe II. Figure 7 shows the effects of adding turbulence to our baseline model. The set of four spectra in the upper right corner of the figure shows the effect of increasing microturbulence. For the LOC models shown here, a microturbulent velocity $v_{turb} \geq 100$ km s$^{-1}$ is necessary to obtain both the observed shape and equivalent width of the UV bump in these models which are integrated over the full log $\Phi$ – log $n_H$ plane.

The rest of the figure illustrates the equivalent width and shape for individual cloud models on the log $\Phi$ – log $n_H$ plane. It is seen that there is a fairly broad area on the plane where both the shape and equivalent width criteria are satisfied. However, one cannot just claim that a single cloud with some specific ($\Phi$, $n_H$) that satisfies these criteria concerning Fe II emission is therefore a viable model of the full BELR, because such a cloud would not make all of the other observed emission lines from other elements. This latter point is shown in the numerous contour plots in Korista et al. (1997) showing the emissivity in lines of many ions as functions of $\Phi$ and $n_H$ (even though the Korista et al calculations included only a simplified Fe$^+$ model atom to account for its contribution to the heating and cooling, the results for ions other than Fe II are basically unchanged).

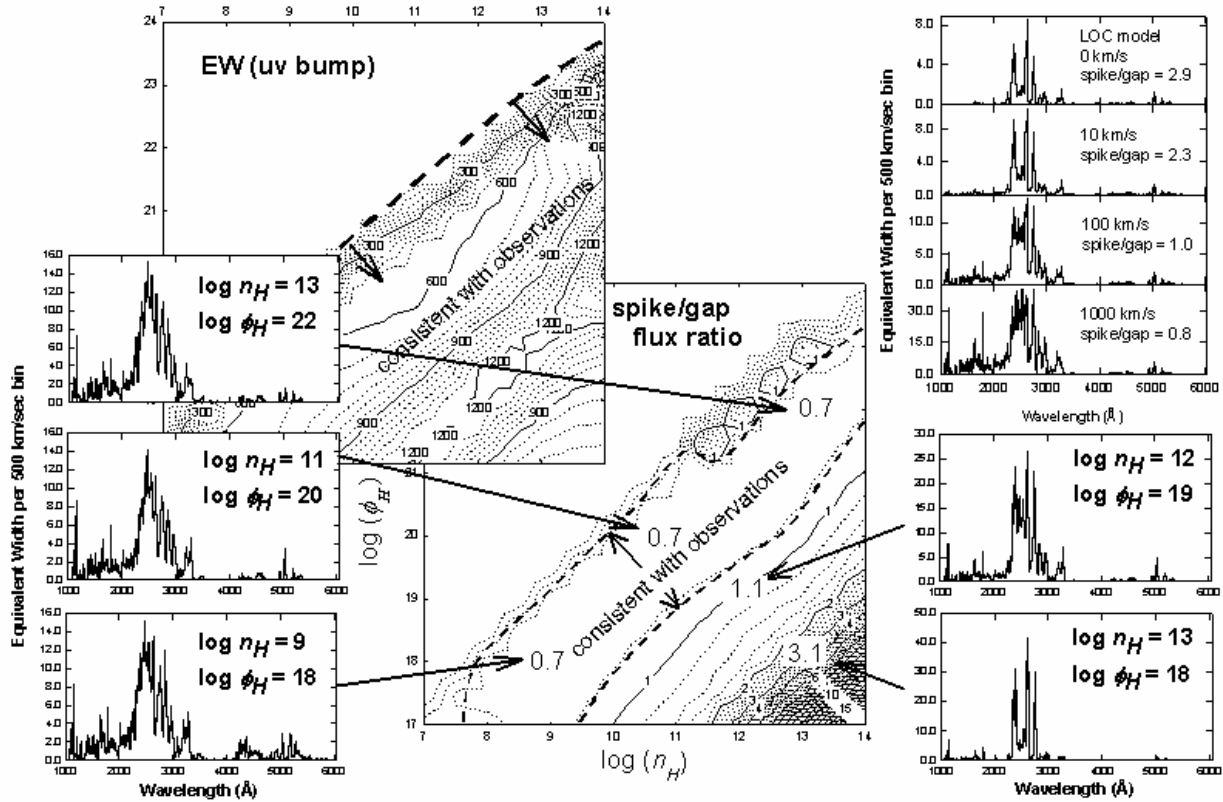

**Figure 7.** (Upper right) Sequence of four Fe II spectra calculated for LOC models having different values of $v_{turb}$, showing how the spikes disappear for higher $v_{turb}$. (Rest of figure) Contour plots of EW (UV bump) and the spike/gap ratio, and some sample Fe II spectra, for $v_{turb}$ = 100 km s$^{-1}$. The large numbers, arrows, etc. have the same meanings as in Figs. 3 and 4. There are now large regions on the contour plots where the models are consistent with the observations.

Models 8 and 9 in Tables 1 and 2 are the results from LOC integrations for $v_{turb}$ = 100 and 1000 km s$^{-1}$. It is seen that these models span a good fit to the observed strengths of the Fe II lines as well as to the strong lines of other elements. We return to discuss possible origins of this local line broadening in §8.

Finally, we note that the highest energy level in our 371-level Fe II atom is 5 eV below the ionization threshold of 16.16 eV. Could the presence of additional levels at higher energies change our predictions with turbulence included? We have already shown (§3) that, without turbulence, the addition of more levels has less than a factor of two effect. But there is more concern with turbulence included because it can enhance the continuum photoexcitation of the upper states. This continuum "pumping" occurs via UV or far-UV transitions whose lower levels are within 1 to 2 eV of ground. Our current model includes thousands of transitions, mainly in the UV, that can pump the atom from these lower levels. Tests in which we turn the pumping off show that in a cloud with $n_H = 10^{12}$ cm$^{-3}$, $\Phi = 10^{20.5}$ cm$^{-2}$ s$^{-1}$, and $N_H = 10^{24}$ cm$^{-2}$, the various forms of photon pumping account for 45 percent of the total flux in the UV bump for $v_{turb}$ = 0 km s$^{-1}$ and 70 percent for $v_{turb}$ = 5 km s$^{-1}$. For a cloud with $n_H = 10^{11}$ cm$^{-3}$, $\Phi = 10^{18}$ cm$^{-2}$ s$^{-1}$, $N_H = 10^{24}$ cm$^{-2}$ and $v_{turb}$ = 5 km s$^{-1}$, the contribution is down to 18 percent. The contribution from photon pumping depends heavily on the exact cloud parameters, but must be very high for the gas with $v_{turb} \geq 100$ km s$^{-1}$ which we suggest here.

Our best estimate of the effect of adding more levels comes from directly comparing our calculations for the 371-level Fe II atom to those for model atoms with higher numbers of levels. Sigut & Pradhan (2003, 2004) and Verner et al. (2003, 2004) have published figures showing the shape of the Fe II UV bump for individual clouds using 827- and 830-level Fe II atoms at several points on the log $\Phi$ – log $n_H$ plane, with $v_{turb}$ values in the range 0-10 km s$^{-1}$. These agree well with the shapes of the Fe II UV bumps predicted by our models having the same or very similar input parameters. In addition, in this very limited sample, all of the models which produce a reasonably high Fe II equivalent width show UV 1, UV 2+3 spikes similar to the ones we find. From this we conclude that our 371-level atom is adequately describing the basic observable results.

## 7. Collisionally Excited Models

We also investigated an alternative concept of the emission line region, in which the gas is collisionally ionized rather than photoionized. Models of this sort, first suggested by Grandi (1981, 1982), have long been championed by Collin-Souffrin and her co-workers (e.g. Joly 1987; Dumont et al. 1998) as the source of emission from Fe II and other low-ionization species, and have also been explored by Kwan et al. (1995). Those papers considered gas at fairly low temperatures, with $T_e$ ~ 5000–8000K, but were concerned primarily with optical Fe II emission.

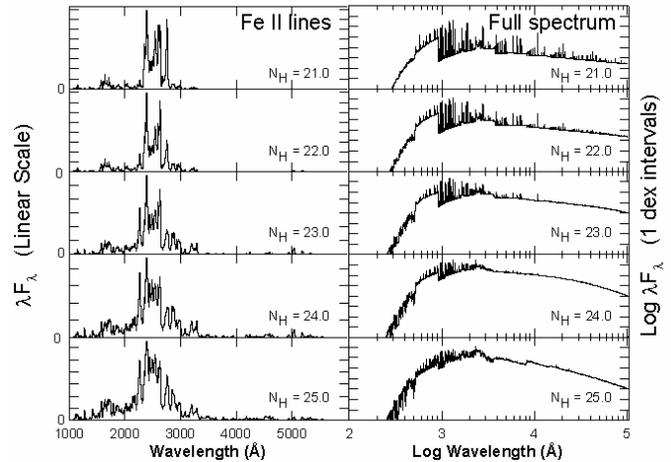

**Figure 8.** Sequence of collisionally excited models with different column densities. This figure shows only every fourth model in our sequence with increasing column density.

Here we explore a broad range of temperatures, calculating series of collisionally excited models for gas with temperatures in logarithmically spaced steps over the range $4000 \leq T_e \leq 250{,}000$K, and densities $n_H = 10^{10}$, $10^{12}$, $10^{14}$ and $10^{16}$ cm$^{-3}$. The line broadening is strictly thermal, and collisions by thermal electrons are assumed to be the only source of ionization. Equivalently, the gas is assumed to be either located in a region that is shielded from the ionizing continuum, or that is sufficiently distant from the continuum source, that photoionization is negligible.

Figure 8 shows a sequence of such models with $T_e = 25{,}250$K and $n_H = 10^{12}$ cm$^{-3}$, for which we have varied the column density over a wide range. The left-hand panels show on a linear scale just the Fe II emission in the UV-optical range, while the right-hand panels show on a log-log scale the full emitted spectrum (including both line and continuum emission from all elements and ionization states that are in Cloudy's database) over a very wide wavelength range. The strong continuum present in the right column of Figure 8 is thermal emission, mainly free-bound and free-free emission, produced by the collisionally ionized gas. Its shape is strongly reminiscent of the "Big Blue Bump", although in all of our models it is too weak to be a major contributor to this observed continuum component.

Large column densities are required to produce Fe II emission similar to what is observed. It is clear from the left-hand panels that column density $N_H > 10^{23}$ cm$^{-2}$ is required to smear out the spikes corresponding to UV 1 and UV 2+3. The right-hand panels show that similarly high column densities are required to suppress emission lines from other elements. On the other hand, for an infinitely large column density a black body spectrum with no emission lines at all would be emitted. Since the calculations take longer and longer at higher column densities, we adopt $N_H = 10^{25}$ cm$^{-2}$ as our representative column density.

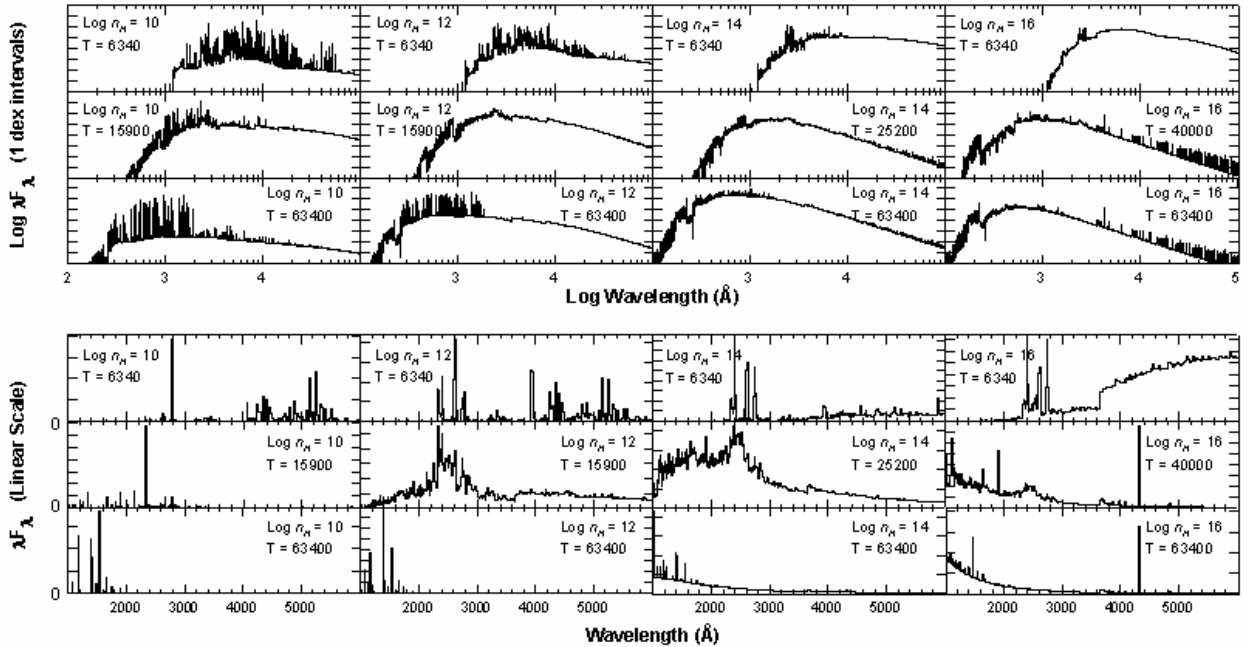

**Figure 9.** Collisionally ionized models, showing the full emitted spectrum including continuum and lines from all ions included in Cloudy. For each of four densities, we show an example that was too cool to produce an Fe II UV bump, one that does produce the bump, and one that is too hot. The upper set of panels shows the full spectrum from 100Å to $10^5$Å on a log-log scale. The lower set of panels shows just the optical-UV region on a linear scale. We computed a grid of 40 such models to find the range over which a strong Fe II UV bump can be produced by a collisionally ionized gas.

Figure 9 shows examples from grids of models having this $N_H$, for the range of temperatures and the four different gas densities mentioned above. Note that in this figure, the lower set of panels now shows the full emitted spectrum, rather than just the Fe II lines. This is because we are interested in knowing about predicted strong emission lines from ions other than Fe II whose absence in the observed spectrum would rule out emission from such collisionally-excited gas as a source of the Fe II UV bump.

It is clear from this grid that for densities $n_H \geq 10^{12}$ cm$^{-3}$, collisionally excited gas with temperatures in the range $10{,}000 \leq T_e \leq 40{,}000$K can produce an emitted spectrum which, over the UV-optical wavelength range, consists of mostly the Fe II UV bump plus an underlying continuum. Our calculations indicate that gas cooler than this would emit other low-ionization lines that are not observed, that the Fe II emission would occur primarily in the optical passband rather than through the UV bump, and that what little FeII UV bump is emitted would be dominated by the UV 1, 2 and 3 spikes rather than a broad bump.

But over the warmer temperature range, the cloud emits mainly Fe II and various continua. We found this result to be very surprising. We had expected the cooling to come out in a much wider variety of emission lines. Fe II emission dominates the UV-optical emission line spectrum due to the complexity of its atomic structure. A gas with this higher density and column density is optically thick in nearly all lines. The lines are strongly thermalized and radiate near their black body limit. $Fe^+$ has many thousands of lines, each radiating near this limit. Other lines that might be expected to be strong, such as Mg II, Na I, or Ca II, are also near this limit, but have only a few transitions, so do not create the same prominent emission as Fe II with its thousands of transitions spread over a range of wavelength. Lines that are normally weak, especially subordinate lines of Fe II, are strong for this reason – they too are near their blackbody limit.

This has one important consequence which we do not pursue. Other ions with complex structure, like Ni II, coming from elements with lower abundance, will create similar features if they too are near their black body limit. But computing atomic data for such complex ions is itself a "Grand Challenge" project (Burke et al. 2002), and rates are not currently available.

## 8. Discussion

We have developed two models that are capable of reproducing the observed shape and intensity of the UV Fe II emission. These are discussed further below.

## 8.1 "Microturbulence" in the BELR

We have confirmed previous results showing that if the Fe II emission is produced in a photoionized gas, significant microturblent velocities must be present; we find $v_{turb} \geq 100$ km s$^{-1}$ for the LOC models considered here. Traditionally, microturbulence is thought of in terms of random motions within the gas. One such mechanism that could lead to such a situation in BELR clouds is magnetohydrodynamic waves in magnetically confined clouds (Rees 1987; Bottorff & Ferland 2000).

However, in our models microturbulence corresponds to any situation in which lines are shifted by $v_{turb}$ over a mean free path of a continuum photon. The observable effects on the emission-line ratios could be due to velocity gradients within smooth flows. Several BELR models include large-scale organized flows involving gas that is radiatively driven from the surface of an accretion disk. (e.g. König & Kartje 1994; Murray et al. 1995; Murray & Chiang 1998; Everett 2002). The hydromagnetic wind model of Blandford & Payne (1982) and Emmering, Blandford & Shlosman (1992) implies a similar smooth gas flow, even though the predicted spectrum (Bottorff et al. 1997) was worked out using computations for individual clouds. The effective microturbulent velocity in such flows, due to the radial shear within the wind or Keplerian shear and local microturbulence within the disk, is expected to be at least several times the local sound speed in the region where Fe II is expected to form (Murray & Chiang 1997, 1998; J. Everett private communication). Thus, the $v_{turb}$ ~100 km s$^{-1}$ or so that is required to produce the observed Fe II UV bump may be present in these windy disk environments.

Other observational evidence favors a turbulent BELR. The cloud model, with only thermal line widths, will produce a line profile with a set of "lumps" corresponding to individual clouds (Capriotti, Foltz, & Byard 1981). Very high signal to noise observations (Arav et al. 1998) show that the profiles are surprisingly smooth in their far wings where the low intensity levels should be due to a fairly small number of clouds at that particular velocity, and that if the clouds have thermal widths, a minimum of $10^7$–$10^8$ clouds is required in the full BELR. The conclusion of Arav et al. is that this suggests that the gas is in a smooth wind. However, individual clouds with turbulent velocities that are much larger than the thermal line widths would be an alternate explanation.

The LOC models used here assume that in addition to the wide radial extent, there is a wide range of gas density at each radial distance, and the parameters used here correspond to summing models with equal weighting at all points over contour plots such as Figures 3, 4, 6 and 7., where the individual cloud models are spaced equally in log $\Phi$ and log $n_H$. Other models (e.g. Rees, Netzer & Ferland 1987; Goad, Obrien & Gondhalekar 1993) make different assumptions about the relation between gas density and radial position, corresponding typically to a weighted averaging of points along various lines on the log $\Phi$ – log $n_H$ plane. As $v_{turb}$ is increased, a larger and larger area of the log $\Phi$ – log $n_H$ plane contains acceptable models of individual clouds. It is possible that for values of $v_{turb}$ less than 100 km s$^{-1}$ there are ways to add up clouds that would satisfy all of the observations listed in tables 1 and 2. We have not attempted to explore all possible BELR models. However, in the cases where there are *no* individual cloud models anywhere on the log $\Phi$ – log $n_H$ plane that simultaneously produce sufficient Fe II equivalent width and a UV bump with the correct shape, we do not see how just adding up some different subset of models can lead to an acceptable result. So our results would seem to rule out non-turbulent photoionized models in a quite general way.

The LOC models approximate the BELR gas as a large number of individual clouds scattered widely across the log $\Phi$ – log $n_H$ plane. We do not try to place constraints such as continuity equations for hydrodynamic flows, and we do not consider the possibility of partial absorption of the ionizing continuum by clouds closer to the center. But the LOC models will still be approximately correct for many imaginable wind situations in which the gas is not heavily shielded by other gas.

## 8.2 A Collisionally Ionized BELR Component?

The alternative model suggested here, a collisionally ionized component of the BELR, has the property that almost the *only* emission lines from it would be the Fe II lines seen in the ultraviolet. This seems like a major drawback, adding yet another specific region to a vague but already complex model of the central part of an AGN.

We note first that this collisionally ionized component does not correspond to a classical accretion disk. The problem is that the column densities through the usual accretion disk models are of order $N_H \sim 10^{26}$ cm$^{-2}$ (Frank, King & Raine 2002; Collin 2003). For such a high column density, along with our constant temperature assumption, the gas would emit only continuum black-body radiation. A real object with an internal heat source would have a temperature decline near the surface. This is why accretion disk models usually treat the disk as a stellar atmosphere, rather than as the semi-transparent collisionally-heated gas cloud described here. The collisionally ionized gas that we are describing here cannot be the gas that makes up the types of accretion disk that are normally discussed in the context of AGN.

Table 3
Equivalent Width[*] of Fe II UV Bump from Collisionally Excited Gas

| Temperature | Log Density (cm$^{-3}$) | | | |
|---|---|---|---|---|
| | 10 | 12 | 14 | 16 |
| 10047 | 2,348 | 889 | 271 | 199 |
| 15924 | 1,338 | 973 | 297 | 278 |
| 25238 | 1,512 | 1,446 | 429 | 242 |
| 40000 | 521 | 108 | 385 | 293 |

[*] Relative to thermal continuum level at 2392Å.

Table 4
Radius[*] of a Collisionally Excited Disk having Necessary Surface Area to produce UV Bump Flux Observed from I Zw 1

| Temperature | Log Density (cm$^{-3}$) | | | |
|---|---|---|---|---|
| | 10 | 12 | 14 | 16 |
| 10047 | 22.3 | 5.1 | 2.9 | 3.5 |
| 15924 | 17.4 | 2.6 | 1.1 | 1.0 |
| 25238 | 12.8 | 1.5 | 0.5 | 0.5 |
| 40000 | 16.9 | 4.2 | 0.4 | 0.3 |

[*] In units of light days

Tables 3 and 4 list some calculated properties of the subset of collisionally ionized models that produce Fe II UV bumps similar in shape to the observed ones. The equivalent width of the bump must be large enough that the bump would be visible against the underlying continuum. Table 3 shows equivalent widths relative to *just* the continuum radiation emitted by the collisionally heated gas. All of the models listed in Table 3 produce more than enough Fe II UV bump radiation relative to the level of the underlying continuum emitted by the same gas. The continuum contribution from this component would be several to many times weaker than the observed continuum radiation seen in Figure 1.

An additional constraint is that the collisionally ionized gas should be able to produce the observed Fe II luminosity without requiring an unreasonably large surface area. Table 4 shows the radius (in light days) of the disk having the surface area needed to produce, from one of its sides, the Fe II UV bump luminosity observed in I Zw 1. For this exercise, we took the luminosity distance to I Zw 1 to be 270 Mpc, so that the observed flux in the Fe II UV bump corresponds to a luminosity of $1.4 \times 10^{43}$ erg s$^{-1}$. The size parameter is presented as a radius to enable easier comparison to the size scales of other structures discussed for the central regions of AGN. In fact, the computed radii are of a reasonable size in comparison to the light-hours to light-days diameters of the variable continuum sources in typical low-luminosity AGN, or as compared to the light-days to light-weeks diameters of the BELR in such objects. Our conclusion is that a collisionally ionized component is a viable model as the source of the Fe II UV bump. We do not know *what* this component would represent physically, but at least it does not need to be of an absurdly large size.

However, a serious drawback of this model is that it requires adding yet another arbitrary component into the complicated zoo of regions and structures within the AGN. For this reason, we favor the model involving microturbulence.

## 8.3 What can we learn from variability?

Do existing observations determine whether the Fe II comes from a highly turbulent BELR or from the collisionally ionized component discussed above? We examine here the possibility that most Fe II emission originates in the collisionally excited component, with the baseline BELR contributing the other bright lines. There would still be a weak Fe II contribution, mainly in the UV 1 and UV 2+3 spikes, from the BELR gas. In AGN with a variable ionizing continuum, we would then expect to see a weak, variable Fe II component dominated by these narrow spikes, superimposed on a stronger, smooth UV bump. The collisionally ionized component would not vary in response to continuum variations.

We searched for such behavior in archival IUE spectra of NGC 5548. This is the Seyfert galaxy with the best studied reverberation behavior. Maoz et al. (1993) found ±20 percent variations in the Fe II UV bump strength. We used the same data, taken from the AGN Watch consortium website[3], to form composite spectra of NGC 5548 when it was in its high and low states.

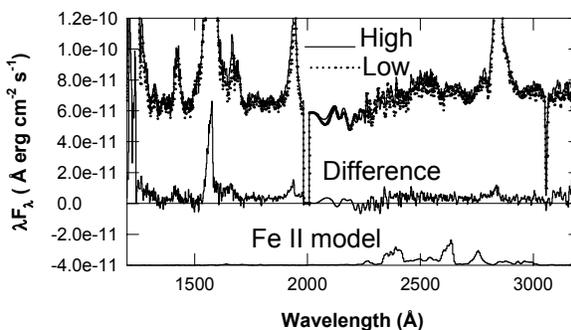

Figure 10 shows the high- and low-state spectra, which fall almost on top of each other. The difference between these two spectra is the spectrum of the variable component. It is also shown in Figure 10, and compared to the Fe II spectrum from our baseline (non-turbulent) photoionized model (§4). There is no sign of the UV 1 or UV 2+3 spikes in the difference spectrum. This suggests that the variable component is from a turbulent photoionized region.

**Figure 10.** (upper panel) High, low and difference spectra for NGC 5548, in the vicinity of the Fe II UV bump. (lower panel) Fe II spectrum calculated for our baseline photoionized model, which has no microturbulence. Note the strong spikes in the model spectrum at $\lambda \sim 2400$ and 2600Å, which do not appear in the difference spectrum.

---

[3] http://www-astronomy.mps.ohio-state.edu/~agnwatch/

However, Fe II is not very strong in the spectrum of NGC 5548. It is notable that in this object, C III] λ1909 and Mg II λ2800, and the Fe II UV bump all vary by only a small amount. The signal:noise ratio in the UV bump in our difference spectrum is quite low. A better case that we hope to study in the future is Akn 120, which is a known variable AGN with quite strong Fe II emission at least in the optical passband, and for which a sequence of IUE spectra suitable for reverberation analysis have been taken (Peterson & Gaskell 1991). It would then be possible to reach a tighter conclusion about this issue.

## 8.4. The Problem of the Optical Fe II Lines

| Table 5 Intensity ratios involving Fe II UV bump, Fe II optical lines, Lyα and Hβ. (measured in the same way in all cases) | | | | | |
|---|---|---|---|---|---|
| | bump/opt | Hβ/opt | Hβ/bump | Lyα/bump | Lyα/Hβ |
| **Observed:** | | | | | |
| I Zw 1 | 1.2 | 0.39 | 0.33 | 2.3 | 7.2 |
| LBQS composite | 1.8 | 1.3 | 0.72 | 5.5 | 7.6 |
| **Models:** | | | | | |
| (1) Baseline log $N_H$ = 23 | 21.4 | 7.1 | 0.33 | 4.6 | 13.8 |
| (2) Baseline log $N_H$ = 24 | 10.9 | 2.8 | 0.26 | 4.4 | 16.9 |
| (3) Baseline log $N_H$ = 25 | 13.1 | 5.6 | 0.43 | 3.6 | 8.5 |
| (4) CNOFe enhanced | 8.7 | 0.7 | 0.08 | 1.5 | 17.5 |
| (5) [Fe/O] = -0.8 | 13.4 | 10.8 | 0.80 | 13.0 | 16.2 |
| (6) [CNO/H] = -0.8, [Fe/O] = -0.8 | 17.0 | 39.1 | 2.30 | 38.4 | 16.7 |
| (7) Power law continuum | 5.4 | 0.7 | 0.12 | 1.6 | 12.7 |
| (8) $v_{turb}$ = 100 km s$^{-1}$ | 45.8 | 8.9 | 0.19 | 2.4 | 12.1 |
| (9) $v_{turb}$ = 1000 km s$^{-1}$ | 54.8 | 5.6 | 0.10 | 1.3 | 13.1 |

None of our photoionized models correctly reproduce the observed strength of the optical Fe II lines. This is shown in Table 5, where we compare for the photoionized models the ratio of the integrated optical line intensity to the UV bump intensity. For this calculation, we used the optical Fe II strength summed over the two wavelength ranges 4450–4750Å and 5080–5460Å, since these are the wavelength ranges for which it is easy to measure the Fe II strength in observed spectra such as those in Figure 1. We have taken the Lyα/Hβ ratio for I Zw 1 from Laor et al. (1997). The strength of Lyα for the LBQS composite was measured by us from the unpublished final LBQS composite spectrum kindly provided by Simon Morris (see Brotherton et al. 2001 for a discussion of this spectrum and its comparison to the published Francis et al. 1991 spectrum). We note that the Lyα emission line does not appear in the z ~ 0.7 subcomposite spectrum, and so the Lyα/Hβ intensity ratio comes from a somewhat different sample. The observed Fe II UV/optical ratio (Table 5, column 2) is 1–2, while the models predict much higher values (reaching 40–50 for the microturbulent models which we favor here).

We do not see how this can be due to reddening, since according to the recent work of Gaskell et al. (2003) even the most highly reddened AGN (which are the lowest luminosity objects) would have only a factor four increase in the observed Fe II UV/optical ratio, and high luminosity objects such as the LBQS quasars are found to have very little reddening. In addition, while the origin of the low Lyα/Hβ ratios observed in AGN is far from a solved problem, the observed ratios in Table 5 are only 2-4 times smaller than the ratios from the photoionized models. The different columns in Table 5 indicate that the strengths of the optical Fe II lines predicted by our models are much weaker than is observed, while the calculated Fe II UV bumps have strengths relative to Lyα and Hβ that are much closer to the observed values.

Figure 11 shows contours of where the optical and UV Fe II lines are actually produced on the log $\Phi$ – log $n_H$ plane. Most of the optical Fe II emission comes from the low $\Phi$, low $n_H$ corner of the diagram, while the UV bump (and Hβ) are produced by gas with a much wider range in $\Phi$ and $n_H$. Nowhere on the log $\Phi$ – log $n_H$ plane does the (UV bump)/(Fe II optical) ratio fall to the observed range of 1-2.

The Fe II UV/optical ratios predicted by our models are within a factor of two of those predicted by Verner et al (2003, 2004) and by Sigut & Pradhan (2003), when we compare single cloud models that have the same input parameters. All of these calculations produce a UV/optical ratio that is significantly higher than the observed ratio of 1.2–1.8.

All the models predict that the optical Fe II emission is not an important energy sink for the clouds, while in the observed spectrum this emission is a major energy sink. This suggests that the bulk of the optical Fe II emission comes from a different region than the UV Fe II emission.

Véron-Cetty, Joly & Véron (2004) have shown that the optical Fe II emission in I Zw 1 is dominated by two systems; one a broader-lined, denser system emitting permitted lines, the other a very low-excitation system that includes many forbidden [Fe II] lines. In Véron-Cetty et al's decomposition of the observed spectrum, the broader-lined component accounts for about 2/3 and the low-excitation about 1/3 of the optical Fe II flux. Thus, either component is still far stronger than is expected from our simulations. Observations comparing the reverberation timescales for optical and UV Fe II lines may help to better discern whether or not these lines come from a common region.

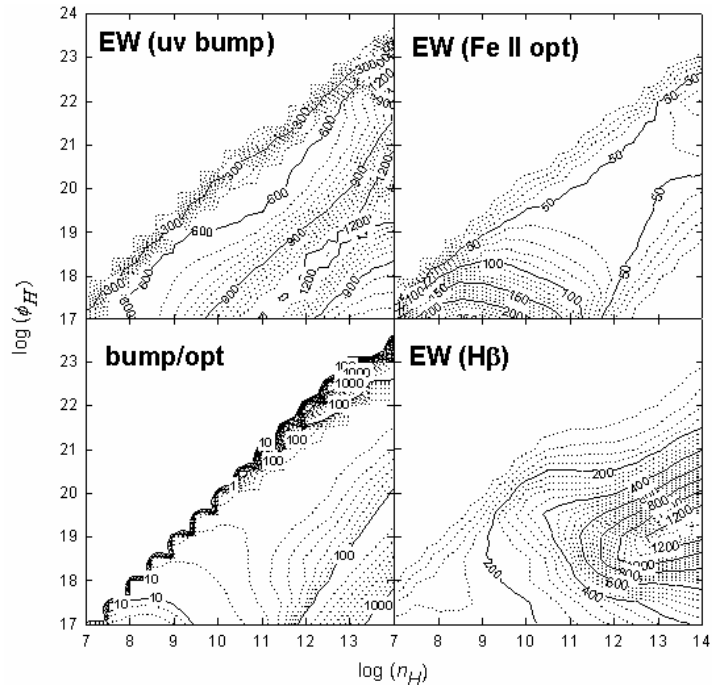

**Figure 11**. Contours of equivalent width as a function of $\Phi_H$ and $n_H$ for (top-left) the Fe II UV bump, (top-right) the Fe II optical emission as defined in the text, and (bottom-right) Hβ. The bottom-left panel shows the ratio of the Fe II (UV bump)/optical equivalent widths. Note that the contours in the bottom-left panel are logarithmically spaced, while the contours are spaced linearly in the other three panels. These plots are for the photoionized models with $v_{turb} = 100$ km s$^{-1}$. Equivalent widths are relative to the continuum level at 2400Å (Fe II UV bump) or 4861Å (Fe II opt, Hβ).

The situation for the collisionally ionized models is better, but still far from perfect. Figure 9 shows that for some of the lower temperature, lower density cases these models produce strong optical Fe II lines but little other UV-optical line emission (except for Mg II 2800). Any optical/UV Fe II line ratio can then be produced just by choosing a correct ratio of surface areas for the gas emitting at low and high temperatures. However, Figure 12 shows that the optical emission from such low-temperature models is not really a good match to that observed from I Zw 1. Low density gas produces most of its Fe II emission in the optical bands, but the observed multiplets 37 and 38 are missing in the model spectrum, while the model produces strong forbidden-line spikes near λ5172Å and λ5280Å which are not observed. Higher density models with $n_H = 10^{14}$ or $10^{16}$ cm$^{-3}$ produce optical Fe II emission with relative strengths much more like what is observed, but also produce UV emission lines vastly stronger than is observed, and which emit in the form of strong UV 1 and UV 2+3 spikes. In addition, producing the optical Fe II lines from a low temperature collisionally ionized component would require still another new component in the BELR.

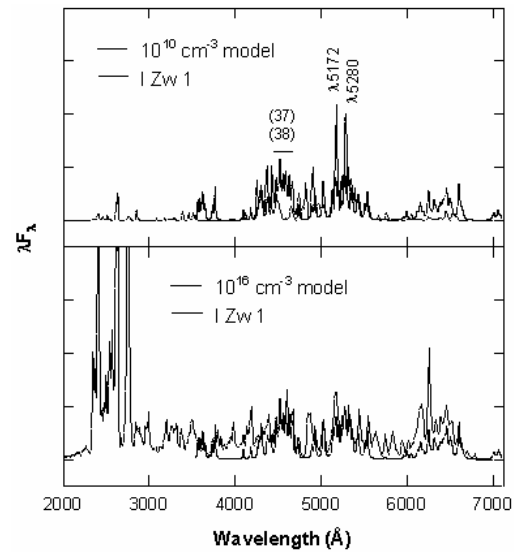

**Figure 12.** Comparison of predicted Fe II optical emission spectrum from a collisionally ionized gas with $T_e = 6340$ K and two different densities, to the template Fe II spectrum derived by Véron-Cetty et al. (2004).

We plan to return to an investigation of the optical Fe II lines at a future date, but for now the observed strengths of these features are not explained by our models.

## 8.5 The Iron Abundance in AGN

We started this work with the goal of quantitatively measuring the degree of Fe enrichment in AGN, as a means of dating the onset of Type Ia supernovae in young galaxies. Our results show that this is more difficult than had been hoped. The problem is that while the relative strength of Fe does depend somewhat on the Fe abundance (our models 4, 5 and 6), it also depends sensitively on other parameters: the turbulent velocity in the case of photoionized gas, or the relative surface area in the case of collisionally ionized gas. We know from the shape of the UV bump that one of these latter two factors definitely affects the strength of the Fe II emission.

Our preferred model for explaining the observed UV properties of the Fe II emission is that it comes from photoionized gas with large velocity gradients. Verner et al. (2003; 2004) have suggested that the Fe II optical line strengths might offer a better measurement of the iron abundance in such a situation, and our results support that idea. As already noted (§6), the overall energy "emitted" by FeII increases with increasing microturbulence, as the optically thick transitions emit more efficiently and especially as photon pumping becomes more dominant (Figure 7). However, Figure 7 also shows that it is the far-UV transitions (1000 – 2000Å) that increase by the largest fraction with increasing microturbulence, followed by the UV transitions (2000 – 3000Å), and lastly the optical transitions (4000 – 6000Å). Consequently, we expect the optical transitions to be more sensitive than the UV transitions to the iron abundance in the presence of uncertain significant microturbulence or other local extra-thermal gas motions.

However, a basic problem discussed in §8.4 is that none of the existing models correctly predict the strengths of the Fe II optical lines. In addition, the Verner et al. (2003, 2004) models are for individual gas clouds, many of which do not satisfactorily reproduce the observed Fe II equivalent widths or the strengths of emission lines from many other elements. Any successful model must at least roughly fit all of the observations. For these reasons we do not think that it is yet possible to use the optical lines to measure the Fe abundance in AGN.

A different route might be to use other line ratios to measure the microtrbulent velocity, so that the Fe II UV bump could then still be used to measure the Fe abundance. For example, the strong sensitivity to photon pumping of the UV transitions near 1800Å (see Figure 7), including the set of transitions associated with relatively isolated multiplet UV 191 near 1787Å, should be diagnostic of the presence of significant local extra-thermal line broadening. In fact Baldwin et al. (1996, their Appendix C) suggested and Bottorff et al. (2000) showed that the presence of significant microturblent velocities ($v_{turb} > 10$ km s$^{-1}$) within the FeII emitting gas is accompanied by relatively strong emission from UV 191 1787Å emission as well as that from SiII 1263, 1307, 1808. The flux ratios Si II 1263/Si II 1808, Si II 1307/Si II 1808, and Fe II 1787/Si II 1808 all increase with increasing microturbulence velocity. They also found that C II 1335 and Al III 1860 are also sensitive to the local line width. We hope to explore this approach in a future paper.

Differences in the microturblent velocity emerge as a likely cause of the observed broad range of Fe II emission properties. Large microturbulence could correspond to either increased MHD wave motion, or to a stronger wind acceleration. In the wind picture, the shear across a photon's mean free path is related to the acceleration, so the larger "microturbulence" in this picture might correspond to larger Eddington ratio, or other changes in the central engine. Observations might be able to tell whether the range of Fe II emission reflect a range in the turbulence, because other lines also change with increasing microturbulence (Bottorff et al. 2000), and large emission line data bases could be searched for independent measures of turbulence. In this picture Fe II-quiet objects would have small turbulent line widths, which might also be revealed by high signal to noise observations like those of Arav et al. (1998). Fe II emission also correlates with other properties, such as radio luminosity and eigenvector 1. Could these influence, or be influenced by, the turbulence?

## 9. Conclusions

We investigated the formation of Fe II emission lines in AGN. We used a 371-level model of the $Fe^+$ ion which produces sufficient accuracy for comparisons to the observed spectra, and which runs rapidly enough that we can calculate Broad Emission Line Region (BELR) cloud models spanning a large range of parameter space. We have used this capability to search for BELR models which reproduce not only the observed Fe II emission properties, but also the relative strengths of the strong emission lines of other elements (Lyα, C IV 1549, Mg II 2800, Hβ, etc.) in a situation where the emission comes from gas spanning a wide range in radial distance from a central ionizing source (as reverberation observations show must be the case). In this way, our models are far more realistic than calculations of the emitted spectrum of a single cloud.

We have shown that a baseline, photoionized model with no microturbulence cannot simultaneously reproduce the observed shape and equivalent width of the Fe II UV bump (the broad feature between 2200 and 2800Å), for any choice of ionizing photon flux and gas density. The shape problem takes the form of the models emitting too much of their energy through the Fe II UV 1, 2 and 3 multiplets, which then appear as two strong, narrow spikes in the spectrum instead of the observed smooth UV bump. High Fe abundances and large column densities *cannot* produce a UV bump that simultaneously has the observed shape and equivalent width. The basic problem is that the energy added by photoionization is deposited at shallow depths into the cloud, depths where the cooling can occur through the UV resonance lines.

The only parameter we could find that leads to acceptable photoionized models is to add microturbulence with $v_{turb} \geq 100$ km s$^{-1}$. As was first suggested by Netzer & Wills, this has the effect of giving the higher FeII multiplets access to a much greater range of exciting continuum photons, while the UV 1, 2 and 3 multiplets are saturated, so that a smooth UV bump is produced. However, this need not indicate true turbulence; it could equally well represent a smooth gas flow in which the velocity changes by $\geq 100$ km s$^{-1}$ over one mean free path of a continuum photon. This latter case is consistent with currently-popular models in which the BELR gas is a wind flowing off a rotating accretion disk.

The one major observational result that this sort of model cannot easily explain is the Fe II optical/UV ratio, which is observed to be considerably higher than is predicted by our (or any other) photoionized models.

We showed that the observed shape and equivalent width of the UV bump can also be reproduced if the Fe II emission comes from a separate collisionally ionized component. We find that gas with temperature $5000 \leq T_e \leq 20,000$K, density $n_H \sim 10^{12}$–$10^{16}$ cm$^{-3}$, and column density $N_H \sim 10^{25}$ cm$^{-2}$ will emit primarily Fe II lines, with the observed shape and with more than adequate equivalent width in the UV bump. Since this gas does not emit strongly in lines of other elements, it would have to constitute a different component than the classic BELR gas. The region cannot have a sufficient column density to represent the sort of accretion disks that are usually suggested as the source of the Big Blue Bump.

Although the line formation theory by itself cannot discriminate between these two scenarios, which represent very different states of the gas in central regions of AGN, there are several observational tests that could determine what is happening. We prefer microturbulence as an explanation, in order to avoid adding yet another arbitrary line-emitting region to an already complicated BELR model. The only advantage we can see for the collisionally excited model is that it could be made to satisfactorily reproduce the observed optical/UV Fe II ratio, by adding separate high- and low-temperature regions.

If either of these scenarios are a correct explanation of the source of the Fe II emission lines, the measurement of the iron abundance from the FeII emission in quasars becomes a more difficult problem. This is because the strength of Fe II emission relative to lines from other elements such as Mg is determined by factors other than the relative abundance: $v_{turb}$ in the microturbulent, photoionized case, and the relative size of the collisionally excited region in the other case.

We are grateful to the referee, Hagai Netzer, and to Mark Bottorff, Gene Capriotti, John Everett, Simon Morris, Norm Murray, Anil Pradhan, Aaron Sigut, Katya Verner and Bev Wills for discussions and for access to their data and calculations. JAB and ALC acknowledge financial support from NSF through grant AST-0305833 and from NASA through grant NAG5-13075. The research of GJF is supported by NSF through grant AST-0307720 and NASA through grants NAG5-8212 and NAG5-12020.